\documentclass[aps]{article}
\usepackage[T1]{fontenc}
\usepackage[latin1]{inputenc}

\makeatletter

\providecommand{\LyX}{L\kern-.1667em\lower.25em\hbox{Y}\kern-.125emX\@}

 \newcommand{\lyxaddress}[1]{
   \par {\raggedright #1 
   \vspace{1.4em}
   \noindent\par}
 }

\makeatother

\begin{document}

\title{Formation of \( d \)-wave superconducting order in a randomly doped lattice}

\author{V. M. Loktev\label{bog} \thanks{
Author to whom correspondence should be addressed. E-mail: vloktev@bitp.kiev.ua
}, Yu. G. Pogorelov\label{cfp}}

\maketitle

\lyxaddress{\ref{bog}Bogolyubov Institute for Theoretical Physics, NAS of Ukraine, Metrologichna
str. 14b, 03143 Kiev, Ukraine; \ref{cfp}CFP and Departamento de Física da Faculdade
de Ciências da Universidade do Porto, Rua do Campo Alegre, 687, 4169-007 Porto,
Portugal}

\begin{abstract}
We consider the interplay between superconducting coupling and dopant impurity
scattering of charge carriers in planar square lattice systems and examine the
physical conditions (doping level, temperature, local symmetry of coupling and
scattering potentials) necessary in this model system to obtain a \( d \)-wave
superconducting order, like that observed in real doped cuprate HTSC materials.
Using the Lifshitz model for the disorder introduced into system by dopants,
we analyze also the non-uniform structure of such \( d \)-wave parameter, including
both its magnitude and phase variation. The results indicate that \( d \)-wave
superconductivity turns possible in a doped metal until it can be destroyed
at too high doping levels.

\end{abstract}

\section{Introduction}

The studies of the effect of impurities and defects on superconducting (SC)
properties of metals (including SC alloys) began practically as early as the
BCS theory had been built. In particular, the classical papers by Anderson \cite{And}
and by Abrikosov and Gor'kov \cite{AbGor} indicated a substantial difference
between magnetic and non-magnetic impurities in superconductors. If an addition
of non-magnetic impurities has practically no effect on the value of transition
critical temperature \( T_{c} \), the presence of spin at impurity atom (leading
to the Kondo effect in a normal metal) results in pair-breaking, that is, it
transforms a singlet Cooper pair into an unstable triplet and rapidly suppresses
\( T_{c} \). In all the known cases, the \( s \)-type, or isotropic, SC order
(apart from the heavy-fermion systems, where it pertains to the \( p \)-type
and \( T_{c} \) is extremely low) and, correspondingly, the isotropic gap near
the Fermi level were considered.

The discovery of high-\( T_{c} \) superconductivity (HTSC) in copper oxides
faced the physicists to a number of problems which still remain a challenge
for the theory. Undoubtedly, this includes the issue of HTSC mechanism, the
strong dependencies of many (both SC and normal) properties of copper oxides
on charge carriers concentration, specifics of weakly doped systems (first of
all, existence of a pseudogap at temperatures above \( T_{c} \)), formation
of stripe structures, etc. (see, e.g., the review articles \cite{Shen},\cite{Lok},\cite{Puch},\cite{LokQuick}).
Such a problem is also presented by the impurity effect on SC properties of
HTSC systems. These differ from {}``old{}'' or {}``conventional{}'' superconductors
not only in higher \( T_{c} \) and \( d \)-wave anisotropy of the order parameter
but also in the fact that here magnetic and non-magnetic impurities change their
roles: the former are weak suppressors for \( T_{c} \) \cite{Bonn}, \cite{Maeda},
while the latter (in particular, Zn substituting Cu in cuprate layers) lead
HTSC to a fast decay \cite{Bonn}, \cite{Fuku}, \cite{Pan}. Many aspects of
impurity effects in superconductors with anisotropic (including \( d \)-wave)
pairing were already theoretically studied in Refs. \cite{Lee},\cite{Bal},\cite{Pog},\cite{Ners},\cite{Ghos},\cite{Durst},
using a range of models and approximations. However, these (and many other)
papers did not include one of the most essential features of HTSC systems, the
fact that they cannot be non-impure.

In other words, most of theoretic approaches to HTSC are based on the concept,
formulated and applied in the pioneering works \cite{And},\cite{AbGor}: one
starts from an ideal (2D or quasi-2D) metal with given Fermi energy \( \varepsilon _{{\rm F}} \),
defined by the density of free carriers, and then consider perturbation of independently
existing SC condensate by some extrinsic (magnetic or non-magnetic) impurities.
Their action, local or global, affects the pre-formed and condensed singlet
pairs. Of course, this formulation is reasonable but it lacks an essential moment
for the conductance in copper oxides: all the HTSC's are doped metals, where
(like the doped semiconductors) each carrier is provided by insertion of a donor
or acceptor into the system. In its turn, this implies that HTSC's are intrinsically
impure systems with an inherent disorder,\footnote{%
Here we don't consider possible formation of stripe structures, where ordered
or disordered distribution of dopants can not be yet confirmed by any reliable
data.
} and the number of impurity ({}``foreign{}'' included) atoms can not be less
than at least the number of charge carriers. If one has in normal metals the
condition \( k_{{\rm F}}\ell \gg 1 \) (\( k_{{\rm F}} \) being the Fermi momentum,
and \( \ell  \) the carrier mean free path between collisions with impurity
atoms) \cite{Mott}, it turns to \( k_{{\rm F}}\ell \sim 1 \) in HTSC's, and
they pertain to the family of \emph{{}``bad{}'' metals} with both \( k_{{\rm F}} \)
and \( \ell  \) defined by the doping.

Perhaps, the first attempt to consider in a self-consistent way the characteristic
tendencies for HTSC, pairing of the carriers and their localization on impurity
atoms, was made in the authors' work \cite{LokPog}. It discussed the phase
diagram of doped 2D metal in presence of \( s \)-wave pairing and showed that
SC is not possible neither at low impurity concentration \( c<c_{0}\sim \varepsilon _{0}/W \)
(when all the carriers are localized near impurities with localization energies
\( \varepsilon _{0} \) much less of the bandwidth \( W \), so that \( c_{0} \)
is typically few percents) nor at too high \( c \) (when the pair inverse lifetime
times \( \hbar  \) exceeds the SC gap). There, in general, the self-consistency
is related either to the SC order parameter (like the common Bardeen-Cooper-Schrieffer
or Bogolyubov-de Gennes treatments) and to the chemical potential. 

The present work is aimed to extend the approach by Ref. \cite{LokPog} to the
case of \( d \)-wave SC coupling and to trace the formation of the corresponding
order parameter. It is motivated, not to the least degree, by an apparent controversy
between the experimental evidence for \( d \)-symmetry of the order parameter
in HTSC \cite{Tsuei}, \cite{Scal}, \cite{Van}, and theoretical claim that
anisotropic pairing should not survive in presence of chaotically distributed
isotropic scatterers \cite{Abr1}. For the sake of simplicity, we restrict the
consideration to the doping range \( c>c_{0} \), where the self-consistency
is only relevant for the SC order parameter, while the chemical potential can
be put\footnote{%
However, it is known that \( \mu  \) can differ essentially from the Fermi
energy \( \varepsilon _{{\rm F}} \) in the limit of very low doping (see, e.g.,
\cite{Lok}, \cite{LokQuick}).
} \( \mu \approx \varepsilon _{{\rm F}}\approx 3cW/4 \). Then, we distinguish
between two types of impurity effects by doping. The first, the so called homogeneous
effects, are displayed by translationally invariant single-particle Green functions
(SPGF). They were studied earlier by various means \cite{WA} but, as a rule,
introducing the disorder through a single parameter \( V_{{\rm A}} \) of Anderson's
model \cite{WA}. In contrary, we employ the Lifshitz' model of disorder \cite{Lif},
characterized by two independent parameters: \( c \) and the impurity potential
\( V_{{\rm L}} \). They produce an equivalent \( V_{{\rm A}}\sim \sqrt{c\left( 1-c\right) }V_{{\rm L}} \),
but not vice versa. Within this model, more adequate for doped HTSC systems,
we conclude about persistence of \( d \)-wave order parameter under homogeneous
impurity effects. Also we explicitly consider other type of effects, inhomogeneous,
due to local variations of order parameter near impurity centers. This involves
\emph{two-particle} Green functions (TPGF), besides usual SPGF, and yields in
possible limitation for SC at high enough dopant concentrations.

At least, we would like to appreciate this great honor and also a pleasant opportunity
for us to publish this work in the Low Temperature Physics issue dedicated to
the memory of outstanding physicist L.V. Shubnikov, whose contribution to the
low temperature physics in general and to superconductivity in particular can
not be overestimated.

\section{Hamiltonian and Green functions}

We start from the model electronic Hamiltonian in band representation\begin{equation}
\label{ham}
H=\sum _{\mathbf{k}}\left\{ \sum _{\sigma }\varepsilon _{\mathbf{k}}c_{\mathbf{k},\sigma }^{\dagger }c_{\mathbf{k},\sigma }-\frac{1}{N}\sum _{\mathbf{k}^{\prime }}\left[ V\gamma _{\mathbf{k}}\gamma _{\mathbf{k}^{\prime }}c_{\mathbf{k},\uparrow }^{\dagger }c_{-\mathbf{k},\downarrow }^{\dagger }c_{-\mathbf{k}^{\prime },\downarrow }c_{\mathbf{k}^{\prime },\uparrow }-\right. \right. 
\end{equation}

\[
\left. \left. -V_{{\rm L}}\sum _{\mathbf{p},\sigma }{\rm e}^{i\left( \mathbf{k}^{\prime }-\mathbf{k}\right) \cdot \mathbf{p}}c_{\mathbf{k}^{\prime },\sigma }^{\dagger }c_{\mathbf{k},\sigma }\right] \right\} ,\]
where \( c_{\mathbf{k},\sigma } \) and \( c_{\mathbf{k},\sigma }^{\dagger } \)are
the Fermi operators for a charge carrier with wavevector \( \mathbf{k} \) and
spin \( \sigma  \). The simplest band energy \( \varepsilon _{\mathbf{k}}=4t-2t\left( \cos ak_{x}+\cos ak_{y}\right)  \),
with full bandwidth \( W=8t \), is expressed through the amplitude \( t \)
of carrier hopping between nearest neighbor sites\footnote{%
So, we do not take into account next neighbor hoppings.
} (of the total \( N \) in the lattice with a constant \( a \)). The parameter
\( V \) models the attraction between two carriers with opposite spins on such
sites, the factor \( \gamma _{\mathbf{k}}=\left( \cos ak_{x}-\cos ak_{y}\right) \theta \left( \varepsilon _{{\rm D}}^{2}-\xi _{\mathbf{k}}^{2}\right)  \)
has the \( d \)-wave symmetry and is effective only for quasiparticle energies
\( \xi _{\mathbf{k}}=\varepsilon _{\mathbf{k}}-\varepsilon _{{\rm F}} \) smaller
than the {}``Debye energy{}'' \( \varepsilon _{{\rm D}} \). The latter is
understood as a characteristic energy of intermediate (Froelich) boson, and
in what follows we suppose the condition \( \varepsilon _{{\rm D}}<\mu  \)
to hold and a BCS-shell to exist (the alternative \( \mu <\varepsilon _{{\rm D}} \),
possible for underdoped HTSC systems, will be considered elsewhere). The impurity
perturbation \( V_{{\rm L}} \) expresses the shift of on-site electronic energy
on a random dopant site \( \mathbf{p} \), where the negative sign takes an
explicit account of carrier attraction to the ionized dopant and, for simplicity,
we consider this perturbation localized on a single site. With usual BCS ansatz:
\( c_{-\mathbf{k},\downarrow }c_{\mathbf{k},\uparrow }=\left\langle c_{-\mathbf{k},\downarrow }c_{\mathbf{k},\uparrow }\right\rangle +\varphi _{\mathbf{k}} \),
and in neglect of quadratic terms in pair fluctuations \( \varphi _{\mathbf{k}} \),
Eq. \ref{ham} leads to a bilinear form for \( H^{\prime }=H-\mu N \): \begin{equation}
\label{bil}
H^{\prime }=\sum _{\mathbf{k}}\left[ \sum _{\sigma }\xi _{\mathbf{k}}c_{\mathbf{k},\sigma }^{\dagger }c_{\mathbf{k},\sigma }-\left( \Delta _{\mathbf{k}}c_{-\mathbf{k},\downarrow }c_{\mathbf{k},\uparrow }+h.c.\right) -\right. 
\end{equation}
\[
\left. -\frac{V_{{\rm L}}}{N}\sum _{\mathbf{p},\mathbf{k}^{\prime },\sigma }\rm e^{-i\left( \mathbf{k}^{\prime }-\mathbf{k}\right) \cdot \mathbf{p}}c_{\mathbf{k}^{\prime },\sigma }^{\dagger }c_{\mathbf{k},\sigma }\right] .\]
 Here the gap function is defined by the self-consistency relation\begin{equation}
\label{gap}
\Delta _{\mathbf{k}}=\frac{V\gamma _{\mathbf{k}}}{N}\sum _{\mathbf{k}^{\prime }}\gamma _{_{\mathbf{k}\prime }}\left\langle c_{\mathbf{k}^{\prime },\uparrow }c_{-\mathbf{k}^{\prime },\downarrow }\right\rangle ,
\end{equation}
 extending the common BCS gap equation to the \( d \)-wave case. A non-uniform
system can be treated within the formalism used formerly for impurity problems
in SC \cite{Pog},\cite{LokPog}, passing to the Nambu spinors \( \psi _{\mathbf{k}}^{\dagger }=\left( c_{\mathbf{k},\uparrow }^{\dagger },c_{-\mathbf{k},\downarrow }\right)  \)
and \( \psi _{\mathbf{k}} \), and defining the Fourier transformed matrix Green
function (GF) \begin{equation}
\label{gf}
\widehat{G}_{\mathbf{k},\mathbf{k}^{\prime }}\left( \varepsilon \right) \equiv \ll \psi _{\mathbf{k}}|\psi _{\mathbf{k}^{\prime }}^{\dagger }\gg _{\varepsilon }=\int _{-\infty }^{0}{\rm e}^{i\left( \varepsilon -i0\right) t}\left\langle \left\{ \psi _{\mathbf{k}}\left( t\right) ,\psi _{\mathbf{k}^{\prime }}^{\dagger }\right\} \right\rangle dt.
\end{equation}
Here \( \widehat{A} \) denotes a \( 2\times 2 \) matrix in Nambu indices,
\( \left\langle \ldots \right\rangle  \) is the quantum statistical average,
and \( \left\{ a\left( t\right) ,b\left( 0\right) \right\}  \) the anticommutator
of Heisenberg operators. In the GF's below we omit their explicit dependence
on energy \( \varepsilon  \) but distinguish between their diagonal and non-diagonal
forms in Nambu (N) and momentum (M) indices. Then, applying the Heisenberg equation
of motion (EM) \( i\hbar \partial \psi _{\mathbf{k}}/\partial t=\left[ H^{\prime },\psi _{\mathbf{k}}\right]  \)
in Eq. \ref{gf}, we arrive at the Dyson's type EM for SPGF's: \begin{equation}
\label{dys}
\widehat{G}_{\mathbf{k},\mathbf{k}^{\prime }}=\widehat{G}_{\mathbf{k}}^{\left( 0\right) }\delta _{\mathbf{k},\mathbf{k}^{\prime }}-\widehat{G}_{\mathbf{k}}^{\left( 0\right) }\widehat{V}\sum _{\mathbf{p},\mathbf{k}^{\prime \prime }}{\rm e}^{i\left( \mathbf{k}-\mathbf{k}^{\prime \prime }\right) \cdot \mathbf{p}}\widehat{G}_{\mathbf{k}^{\prime \prime },\mathbf{k}^{\prime }}
\end{equation}
where the non-perturbed SPGF \( \widehat{G}_{\mathbf{k}}^{\left( 0\right) }=\left( \varepsilon -\xi _{\mathbf{k}}\widehat{\tau }_{3}-\Delta _{\mathbf{k}}\widehat{\tau }_{1}+i0\right) ^{-1} \),
and the scattering matrix \( \widehat{V}=V_{{\rm L}}\widehat{\tau }_{3} \)
include the Pauli matrices \( \widehat{\tau }_{i} \). 

For a disordered system, the relevant (observable) characteristics are described
by the so-called self-averaging GF's, whose values for all particular realizations
of disorder turn practically non-random, equal to those averaged over disorder
\cite{LifGrPa}. The most important example of such a function is the M-diagonal
SPGF, \( \widehat{G}_{\mathbf{k}}\equiv \widehat{G}_{\mathbf{k},\mathbf{k}} \).The
general solution for Eq. \ref{dys} in this case can be written (see Appendix
E) as \begin{equation}
\label{sol}
\widehat{G}_{\mathbf{k}}=\left\{ \left[ \widehat{G}_{\mathbf{k}}^{\left( 0\right) }\right] ^{-1}-\widehat{\Sigma }_{\mathbf{k}}\right\} ^{-1},
\end{equation}
where the self-energy matrix \( \widehat{\Sigma }_{\mathbf{k}} \) is given
by the so-called fully renormalized group expansion (GE) \cite{Iv}, \cite{ILP},\cite{Pog},\cite{LokPog}

\[
\widehat{\Sigma }_{\mathbf{k}}=-c\widehat{V}\left[ 1+\widehat{G}\widehat{V}\right] ^{-1}\times \]
\begin{equation}
\label{gr}
\times \left\{ 1+c\sum _{\mathbf{n}\neq 0}\left[ \widehat{A}_{0\mathbf{n}}{\rm e}^{-i\mathbf{k}\cdot \mathbf{n}}+\widehat{A}_{0\mathbf{n}}\widehat{A}_{\mathbf{n}0}\right] \left[ 1-\widehat{A}_{0\mathbf{n}}\widehat{A}_{\mathbf{n}0}\right] ^{-1}+\cdots \right\} .
\end{equation}
Here the integrated SPGF matrix \( \widehat{G}=N^{-1}\sum _{\mathbf{k}}\widehat{G}_{\mathbf{k}} \),
and the matrices \( \widehat{A}_{0\mathbf{n}} \) of indirect interaction between
scatterers at sites \( 0 \) and \( \mathbf{n} \) are\begin{equation}
\label{a0n}
\widehat{A}_{0\mathbf{n}}=-\widehat{V}\sum _{\mathbf{k}^{\prime }\neq \mathbf{k}}{\rm e}^{i\mathbf{k}^{\prime }\cdot \mathbf{n}}\widehat{G}_{\mathbf{k}^{\prime }}\left[ 1+\widehat{G}\widehat{V}\right] ^{-1}.
\end{equation}
 The rectriction to \( \mathbf{k}^{\prime }\neq \mathbf{k} \) at single summation
in \( \widehat{A}_{0\mathbf{n}} \) should be complemented by \( \mathbf{k}^{\prime \prime }\neq \mathbf{k},\mathbf{k}^{\prime } \)
at double summation in the product \( \widehat{A}_{0\mathbf{n}}\widehat{A}_{\mathbf{n}0} \),
but such restrictions can be already ignored in \( \widehat{A}_{0\mathbf{n}}\widehat{A}_{0\mathbf{n}}\widehat{A}_{\mathbf{n}0} \)
and higher degree terms \cite{Iv}, resulting from expansion of the r.h.s. of
Eq. \ref{gr}.

Many observable characteristics of SC state follow from GF's, using the spectral
theorem representation\begin{equation}
\label{sp}
\left\langle ab\right\rangle =\int _{-\infty }^{\infty }\frac{d\varepsilon }{{\rm e}^{\beta \left( \varepsilon -\mu \right) }+1}{\rm Im}\ll b|a\gg _{\varepsilon },
\end{equation}
where the chemical potential \( \mu  \) is defined by the overall electronic
concentration\begin{equation}
\label{dens}
c=\frac{1}{N}\sum _{\mathbf{k}}\int _{-\infty }^{\infty }\frac{d\varepsilon }{{\rm e}^{\beta \left( \varepsilon -\mu \right) }+1}{\rm Im\: Tr}\widehat{\tau }_{3}\widehat{G}_{\mathbf{k}}.
\end{equation}
On the other hand, \( c \) is just the concentration of dopant centers: \( c=N^{-1}\sum _{\mathbf{p}}1 \),
which give rise to carrier scattering, and the \emph{carrier} concentration
only gets close to (but never exceeds) \( c \) in the regime of doped metal,
for \( c \) above a certain metallization threshold \( c_{0} \) (for quasi-2D
dispersed \( \varepsilon _{\mathbf{k}} \), it is \( c_{0}\sim \exp \left( -\pi W/4V_{{\rm L}}\right) \ll 1 \)
\cite{LokPog}). At this condition, the self-consistency implied by Eq. \ref{dens}
is not necessary, and a good approximation\footnote{%
This approximation is actually justified by the fact that for \( c>c_{0} \)
the Fermi level \( \varepsilon _{{\rm F}} \) of metallic phase is well higher
than the conduction band edge, and one can hardly suppose existence of local
pairs and related inequality \( \mu <\varepsilon _{{\rm F}} \) at these concentrations.
Therefore, in what follows we do not distinguish between \( \mu  \) and \( \varepsilon _{{\rm F}} \).
} for chemical potential is \( \mu \approx 3cW/4 \) (see Appendix A). Then the
gap equation, Eq. \ref{gap}, presents as\begin{equation}
\label{gap1}
\Delta _{\mathbf{k}}=\frac{V\gamma _{\mathbf{k}}}{2N}\sum _{\mathbf{k}^{\prime }}\gamma _{_{\mathbf{k}^{\prime }}}\int _{-\infty }^{\infty }\frac{d\varepsilon }{{\rm e}^{\beta \left( \varepsilon -\mu \right) }+1}{\rm Im\: Tr}\widehat{\tau }_{1}\widehat{G}_{\mathbf{k}^{\prime }},
\end{equation}
and its solution discussed in Appendix C for the uniform case (\( V_{{\rm L}}=0 \))
is simply \( \Delta _{\mathbf{k}}=\Delta \gamma _{\mathbf{k}} \), with the
ratio \( r=2\Delta /k_{{\rm B}}T_{c} \) being \( {\rm e}^{1/3} \) times the
\( s \)-wave BCS value \( r_{{\rm BCS}}\approx 3.52 \).

Another important self-averaging quantity is the integrated SPGF matrix \( \widehat{G} \)
itself, since the density of states \( \rho \left( \varepsilon \right)  \)
is just\begin{equation}
\label{rho}
\rho \left( \varepsilon \right) =\frac{1}{\pi }{\rm Im\: Tr}\widehat{G}.
\end{equation}
 For a non-perturbed system, \( V_{{\rm L}}\rightarrow 0, \) \( \widehat{G}\rightarrow \widehat{G}^{\left( 0\right) }=N^{-1}\sum _{\mathbf{k}}\widehat{G}^{\left( 0\right) }_{\mathbf{k}} \),
and calculation of the imaginary part of \( \widehat{G}^{\left( 0\right) } \)
within the nodal point approximation (Appendix B) leads to the standard \( d \)-wave
density of states: \begin{equation}
\label{rho0}
\rho \left( \varepsilon \right) \rightarrow \rho ^{\left( 0\right) }\left( \varepsilon \right) =\frac{1}{\pi }{\rm Im\: Tr}\widehat{G}^{\left( 0\right) }=\frac{2\varepsilon \rho _{0}}{\Delta }{\rm arcsin}\left[ {\rm min}\left( 1,\frac{\Delta }{\varepsilon }\right) \right] ,
\end{equation}
 where \( \rho _{0}\approx 4/(\pi W) \) is the normal Fermi density of states
of doped (quasi-2D) metal. Respectively, the real part of \( \widehat{G}^{\left( 0\right) } \)is\begin{equation}
\label{reg}
{\rm Re}\widehat{G}^{\left( 0\right) }=\varepsilon \rho _{0}\left[ \frac{W}{\mu \left( W-\mu \right) }-\frac{\pi }{\Delta }\theta \left( \Delta -\varepsilon \right) {\rm arccosh}\frac{\Delta }{\varepsilon }\right] .
\end{equation}
 Then Eqs. \ref{rho0} and \ref{reg} can be unified into a single analytic
form: \begin{equation}
\label{g0an}
\widehat{G}^{\left( 0\right) }=\varepsilon \rho _{0}\left[ \frac{W}{\mu \left( W-\mu \right) }-\frac{\pi }{\Delta }\left( {\rm arccosh}\frac{\Delta }{\varepsilon }-i\frac{\pi }{2}\right) \right] ,
\end{equation}
 since at \( \varepsilon >\Delta  \) one has \( {\rm arccosh}(\Delta /\varepsilon )=i[\pi /2-{\rm arcsin}(\Delta /\varepsilon )] \),
thus restoring Eq. \ref{rho0}. But it is just the growth of (real) arccosh
term at \( \varepsilon <\Delta  \) that permits existence of a low-energy (\( \varepsilon _{res}\ll \Delta  \))
resonance feature in \( {\rm Re}(1+\widehat{G}\widehat{V})^{-1} \), and hence
in \( \rho \left( \varepsilon \right)  \). Such a resonance was formerly discussed
for a \( d \)-wave SC with low enough concentration \( c \) (so that \( \widehat{G}\approx \widehat{G}^{\left( 0\right) } \))
of {}``foreign{}'' impurities producing strong enough perturbation \( V_{{\rm L}} \)
\cite{Pog}, and it is similar to the known low-frequency resonance by heavy
impurities in acoustic phonon spectra \cite{Kag}. However, in the situation
of our interest here, when both \( V_{{\rm L}} \) and \( c \) are not small,
\( \widehat{G} \) can be essentially modified compared to \( \widehat{G}^{\left( 0\right) } \)
and this is expressed in a very complicate way by Eq. \ref{gr}. To simplify
the task, certain self-consistent procedures, like the CPA method, quite useful
in the theory of normal metals \cite{Ell}, can be employed. Similar approach
was formerly proposed for an \( s \)-wave SC doped system \cite{LokPog}, and
here we begin with the analysis of a self-consistent solution for Eq. \ref{dys}
in the \( d \)-wave case.

\section{Uniform doping effects in self-consistent approach}

If the GE series, Eq. \ref{gr}, is restricted to its first term, the self-energy
matrix \( \widehat{\Sigma }_{\mathbf{k}} \) turns in fact independent of \( \mathbf{k} \):\begin{equation}
\label{sigm}
\widehat{\Sigma }_{\mathbf{k}}\rightarrow \widehat{\Sigma }=-c\widehat{V}\left[ 1+\widehat{G}\widehat{V}\right] ^{-1},
\end{equation}
 and substitution of Eq. \ref{sigm} into Eq. \ref{sol} defines the self-consistent
approximation \( \widehat{G}^{\left( sc\right) }_{\mathbf{k}} \) for M-diagonal
SPGF:\begin{equation}
\label{sc1}
\widehat{G}^{\left( sc\right) }_{\mathbf{k}}=\left\{ \left[ \widehat{G}_{\mathbf{k}}^{\left( 0\right) }\right] ^{-1}-\widehat{\Sigma }^{\left( sc\right) }\right\} ^{-1},
\end{equation}
\begin{equation}
\label{sc2}
\widehat{\Sigma }^{\left( sc\right) }=-c\widehat{V}\left[ 1+\widehat{G}^{\left( sc\right) }\widehat{V}\right] ^{-1},
\end{equation}
\begin{equation}
\label{sc3}
\widehat{G}^{\left( sc\right) }=\frac{1}{N}\sum _{\mathbf{k}}\widehat{G}^{\left( sc\right) }_{\mathbf{k}}.
\end{equation}
To solve this system, we first parametrize the self-energy matrix, Eq. \ref{sc2}:\begin{equation}
\label{param}
\widehat{\Sigma }^{\left( sc\right) }=\Sigma _{0}+\Sigma _{1}\widehat{\tau }_{1}+\Sigma _{3}\widehat{\tau }_{3},
\end{equation}
 \( \Sigma _{i} \) being some complex-valued functions of energy. Then integration
in Eq.\ref{sc3} within the nodal point approximation (Appendix D) results in\begin{equation}
\label{Gsc}
\widehat{G}^{\left( sc\right) }=G_{0}+G_{1}\widehat{\tau }_{1}+G_{3}\widehat{\tau }_{3}
\end{equation}
with the coefficients

\begin{equation}
\label{g0}
G_{0}=\left( \varepsilon -\Sigma _{0}\right) \rho _{0}\left[ -\frac{\pi }{2\Delta }\left( {\rm arccosh}\frac{\Delta +\Sigma _{1}}{\varepsilon -\Sigma _{0}}+{\rm arccosh}\frac{\Delta -\Sigma _{1}}{\varepsilon -\Sigma _{0}}-i\pi \right) +\right. 
\end{equation}
\[
\left. +\frac{W}{\mu \left( W-\mu \right) }\right] ,\]

\begin{equation}
\label{g1}
G_{1}=\Sigma _{1}\rho _{0}\left[ -\frac{2i\pi }{\sqrt{(\varepsilon -\Sigma _{0})^{2}-\left( \Delta +\Sigma _{1}\right) ^{2}}+\sqrt{(\varepsilon -\Sigma _{0})^{2}-\left( \Delta -\Sigma _{1}\right) ^{2}}}+\right. 
\end{equation}
\[
\left. +\frac{W}{\mu \left( W-\mu \right) }\right] ,\]
\begin{equation}
\label{g3}
G_{3}=\rho _{0}\left[ \ln \frac{\mu }{W-\mu }+2\Sigma _{3}\frac{(\varepsilon -\Sigma _{0})^{2}-\Delta ^{2}/3-\Sigma _{1}^{2}}{\varepsilon _{{\rm D}}^{3}}\right] .
\end{equation}
 Substituting Eq. \ref{Gsc} into Eq. \ref{sc2}, we arrive at\begin{equation}
\label{sigma}
\widehat{\Sigma }^{\left( sc\right) }=\frac{cV_{{\rm L}}\left[ V_{{\rm L}}\left( G_{0}+G_{1}\widehat{\tau }_{1}\right) -\left( 1+V_{{\rm L}}G_{3}\right) \widehat{\tau }_{3}\right] }{\left( 1+V_{{\rm L}}G_{3}\right) ^{2}-V_{{\rm L}}^{2}\left( G_{0}^{2}-G_{1}^{2}\right) }.
\end{equation}
Comparing Eqs. \ref{sigma}, and \ref{g0}, \ref{g1}, \ref{g3} with Eq. \ref{param},
we immediately conclude that \( \Sigma _{1}=G_{1}=0 \), or that \( \widehat{\Sigma }^{\left( sc\right) } \)
is in fact \emph{N-diagonal,} which is extremely important. Physically, this
means that (within the self-consistent, linear in \( c \) approximation) the
scattering by dopants does not influence the \( d \)-wave order parameter,
and this can be directly related to the fact that the \( s \)-symmetry of impurity
perturbation \( V_{{\rm L}} \) is orthogonal to the \( d \)-symmetry of SC
pairing \( V \). It also applies to more realistic models of dopant perturbation
in HTSC (e.g. with plaquette- or dumbbell-like anisotropy \cite{ILP1}), as
far as their symmetries do not coincide with that of the order parameter. Complications
arise when they do coincide, as was found for isotropic perturbation on \( s \)-wave
order with all three \( \Sigma _{i} \) being non-zero \cite{LokPog}, hence
the apparently {}``harder{}'' \( d \)-wave system turns in fact {}``easier{}''!

Using the fact that \( \Sigma _{1}=0 \) and the relation \( \cosh (x+i\pi /2)=\sin x \),
Eq. \ref{g0} is brought to a very simple form:\begin{equation}
\label{g0a}
\frac{\Delta }{\varepsilon -\Sigma _{0}}=\sin \left( \alpha -\frac{G_{0}}{\pi \rho _{0}}\frac{\Delta }{\varepsilon -\Sigma _{0}}\right) ,
\end{equation}
with \( \alpha =W\Delta /[\pi \mu \left( W-\mu \right) ]\ll 1 \), while the
same comparison for the two non-zero components of \( \widehat{\Sigma }^{\left( sc\right) } \):
\( \Sigma _{0} \) and \( \Sigma _{3} \), gives: \begin{equation}
\label{sc4}
\Sigma _{0}\left[ \left( 1+V_{{\rm L}}G_{3}\right) ^{2}-V_{{\rm L}}^{2}G_{0}^{2}\right] =cV_{{\rm L}}^{2}G_{0},
\end{equation}
\begin{equation}
\label{sc5}
\Sigma _{3}\left[ \left( 1+V_{{\rm L}}G_{3}\right) ^{2}-V_{{\rm L}}^{2}G_{0}^{2}\right] =-cV_{{\rm L}}\left( 1+V_{{\rm L}}G_{3}\right) .
\end{equation}
 From Eq. \ref{sc5} we estimate \( |\Sigma _{3}|\sim cV_{{\rm L}} \), hence
within the relevant energy region \( \left| \varepsilon \right| \ll \varepsilon _{{\rm D}} \),
the function \( G_{3} \) from Eq. \ref{g3} is reasonably well approximated
by a (negative) constant \( g_{3}=\rho _{0}\ln [\mu /(W-\mu )] \). Then Eq.
\ref{sc4} turns quadratic for \( G_{0} \):\begin{equation}
\label{quad}
G_{0}^{2}+\frac{c}{\Sigma _{0}}G_{0}-\left( \frac{1}{\widetilde{V}}\right) ^{2}=0,
\end{equation}
 with \( \widetilde{V}=V_{{\rm L}}/(1+V_{{\rm L}}g_{3})\approx V_{{\rm L}}\ln (1/c_{0})/\ln (3c/4c_{0}) \).
The system, Eqs. \ref{g0a}, \ref{quad}, fully defines the self-energy \( \Sigma _{0} \)
and other uniform physical properties of a disordered \( d \)-wave system,
and its solution can be found (in principle numerically) within the whole relevant
energy range. It turns especially simple if \( |\Sigma _{0}|\ll c\widetilde{V} \)
(this proves to hold at least for \( \varepsilon \ll \varepsilon _{res} \)),
then the proper solution to Eq. \ref{quad} is: \( G_{0}\approx \Sigma _{0}/c\widetilde{V}^{2} \),
and from Eq. \ref{g0a} we obtain the following equation for a single important
function \( \Sigma _{0} \):\begin{equation}
\label{eq}
\frac{\Delta }{\varepsilon -\Sigma _{0}}=\sin \left( \alpha -\frac{\beta \Sigma _{0}}{\varepsilon -\Sigma _{0}}\right) 
\end{equation}
 with \( \beta =\Delta /[\pi c\widetilde{V}^{2}\rho _{0}]\ll 1 \). It defines
the self-consistent density of states\begin{equation}
\label{scdens}
\rho ^{(sc)}\left( \varepsilon \right) =\frac{1}{\pi }{\rm Im\: Tr}\widehat{G}^{(sc)}\left( \varepsilon \right) =\frac{2{\rm Im}\Sigma _{0}\left( \varepsilon \right) }{\pi c\widetilde{V}^{2}},
\end{equation}
 at lowest energies. This approach results free of infrared logarithmic divergencies,
appearing in the integrals of perturbation theory \cite{GorKal}, and thus permits
to avoid applying heavy field theory methods for white-noise scattering potential
\cite{Ners}, whose adequacy to the case of discrete random dopants is not clear. 

The exact value of the density of states at the very center of the gap, \( \rho \left( 0\right) =\rho \left( \varepsilon \rightarrow 0\right)  \),
is also of a particular interest in view of the known claim about existence
of a non-zero {}``universal{}'' value \( \rho \left( 0\right) \sim c/\ln (1/c) \)
if \( V_{{\rm L}} \) is sufficiently strong \cite{Lee}, \cite{Durst}. However,
we conclude from Eq. \ref{eq} that in the limit \( \varepsilon \rightarrow 0 \):\[
\Sigma _{0}\rightarrow \varepsilon \left[ 1+i\frac{\beta }{\ln \left( 2\Delta /\beta |\varepsilon |\right) }\right] ,\]
 and hence the self-consistent density of states at \( \varepsilon \rightarrow 0 \):\begin{equation}
\label{lim}
\rho ^{(sc)}\left( \varepsilon \right) \rightarrow \frac{4\beta ^{2}\rho ^{(0)}\left( \varepsilon \right) }{\pi \ln \left( 2\Delta /\beta |\varepsilon |\right) },
\end{equation}
 vanishes even faster then the non-perturbed function \( \rho ^{\left( 0\right) }\left( \varepsilon \right)  \),
Eq. \ref{rho0}. This produces a certain narrow {}``quasi-gap{}'' (not to
be confused with the pseudo-gap observed at \( T>T_{c} \) in the underdoped
regime) around the center. For comparison, the estimated \( \rho \left( \varepsilon \right)  \)
from the two first terms of Eq. \ref{gr} tends to zero linearly in \( \varepsilon  \)
with corrections \( \sim \varepsilon ^{2} \) \cite{Pog}, while the field-theoretical
analysis \cite{Ners} predicts \( \rho \left( \varepsilon \right) \rightarrow \varepsilon ^{\omega } \),
with the non-universal exponent being (in our notations) \( \omega =\tanh \ln \sqrt{\pi ^{2}\Delta W/2cV_{{\rm L}}^{2}} \),
that is always \( <1 \) and even not excluding \( <0 \). 

The discrepancy between our results and the before mentioned {}``universal{}''
behavior originates in the improper use of unitary limit \( V_{{\rm L}}\rightarrow \infty  \)
at neglect of the \( 1+V_{{\rm L}}G_{3} \) term in Eq. \ref{sc4}, leading
to the relation \( \Sigma _{0}=-c/G_{0} \). But the true limiting relation
is inverse: \( \Sigma _{0}=c\widetilde{V}^{2}G_{0} \), with \( \widetilde{V} \)
\emph{finite} for \( V_{{\rm L}}\rightarrow \infty  \), and also the unitary
limit fails at any\footnote{%
Unless the special case \( 1/\widetilde{V}=0 \) corresponding to \( c\approx 4c_{0}/3 \),
while the actual consideration is for \( c\gg c_{0} \).
} finite \( V_{{\rm L}} \) when \( \varepsilon \rightarrow 0 \). Finally, the
existing experimental data do not confirm the {}``universal{}'' \( \rho \left( 0\right)  \)
value, but seem to favor the conclusion about existence of a strong low-energy
resonance in \( \rho \left( \varepsilon \right)  \) \cite{Pan}, with possible
quasi-gap at the very center \cite{Cov}, though experimental observations at
such low energies of course need extremely low temperatures.

Note however that the self-consistent treatment of the low-energy resonance,
at \( \varepsilon _{res}\sim \Delta \ln (3c/4c_{0})/(\pi \ln 2\pi ) \) for
the case of {}``own{}'' impurities (Fig. \ref{fig1}), requires already solution
of the full system, Eqs. \ref{g0a}, \ref{quad}, and, in view of a probable
underestimate of this hump (like that in normal systems \cite{ILP},\cite{Ell}),
it should be better described by the exact GE, Eq. \ref{gr}.

The obtained \( \widehat{\Sigma }^{\left( sc\right) } \) can be in principle
directly inserted in Eq. \ref{sc1}, in order to use the resulting \( \widehat{G}^{\left( sc\right) }_{\mathbf{k}} \)
for correction of the gap equation, Eq. \ref{gap1}. However, at important there
quasiparticle energies \( \xi _{\mathbf{k}}\sim \varepsilon _{{\rm D}} \),
renormalization effects are negligible, and thus \( \Delta  \) remains well
approximated by the result of Appendix C.

\section{Non-uniform effects}

The SPGF's, considered in the previous section, describe the uniform self-averaging
characteristics of SC state. The next important question is the behavior of
fluctuations of the order parameter (both its amplitude and phase) in an inhomogeneous
system, which should be closely related to the breakdown of superconductivity
in the overdoped regime. A strong local suppression of \( d \)-wave order close
to a single {}``foreign{}'' impurity was predicted theoretically \cite{Pog}
and observed experimentally \cite{Pan}. In the general case of finite concentration
of scatterers, the local \( d \)-wave order can be characterized by the operator\begin{equation}
\label{omega}
\Omega _{\mathbf{n}}=\frac{V}{N}\sum _{\mathbf{k},\mathbf{k}^{\prime }}\gamma _{\mathbf{k}}{\rm e}^{i\left( \mathbf{k}-\mathbf{k}^{\prime }\right) \cdot \mathbf{n}}c_{-\mathbf{k},\downarrow }c_{\mathbf{k}^{\prime },\uparrow },
\end{equation}
 such that its mean value (generally complex) defines the uniform gap parameter:
\( N^{-1}\sum _{\mathbf{n}}\left\langle \Omega _{\mathbf{n}}\right\rangle =N^{-1}\sum _{\mathbf{k}}\gamma _{\mathbf{k}}\Delta _{\mathbf{k}}=\Delta  \)
\cite{Pog}. On the other hand, it is natural to characterize local fluctuations
of order parameter by the mean squared dispersion of \( \Omega _{\mathbf{n}} \)
(identified with the dispersion of gap parameter):\begin{equation}
\label{disp}
\delta ^{2}=\frac{1}{N}\sum _{\mathbf{n}}\left( \left\langle \Omega _{\mathbf{n}}^{2}\right\rangle -\left\langle \Omega _{\mathbf{n}}\right\rangle ^{2}\right) .
\end{equation}
Using Eq. \ref{omega}, we write \begin{equation}
\label{disp1}
\delta ^{2}=\frac{V^{2}}{N^{3}}\sum _{\mathbf{n}}\sum _{\mathbf{k}_{1},\mathbf{k}_{2},\mathbf{k}_{3},\mathbf{k}_{4}}{\rm e}^{i\left( \mathbf{k}_{1}+\mathbf{k}_{2}-\mathbf{k}_{3}-\mathbf{k}_{4}\right) \cdot \mathbf{n}}\gamma _{\mathbf{k}_{1}}\gamma _{\mathbf{k}_{2}}\left[ \left\langle c_{-\mathbf{k}_{1},\downarrow }c_{\mathbf{k}_{3},\uparrow }c_{-\mathbf{k}_{2},\downarrow }c_{\mathbf{k}_{4},\uparrow }\right\rangle -\right. ,
\end{equation}
\[
\left. -\left\langle c_{-\mathbf{k}_{1},\downarrow }c_{\mathbf{k}_{3},\uparrow }\right\rangle \left\langle c_{-\mathbf{k}_{2},\downarrow }c_{\mathbf{k}_{4},\uparrow }\right\rangle \right] ,\]
or, summing up in \( \mathbf{n} \) to close the sum of 4 momenta, and using
the spectral representation, Eq. \ref{sp}, for \( T=0 \): \begin{equation}
\label{disp2}
\delta ^{2}=\frac{V^{2}}{N^{2}}\sum _{\mathbf{k}_{1},\mathbf{k}_{2},\mathbf{q}}\gamma _{\mathbf{k}_{1}}\gamma _{\mathbf{k}_{2}}\left[ \int _{0}^{\infty }d\varepsilon {\rm Im}\ll c_{-\mathbf{k}_{1},\downarrow }c_{\mathbf{k}_{2},\uparrow }|c_{-\mathbf{k}_{2}+\mathbf{q},\downarrow }c_{\mathbf{k}_{1}-\mathbf{q},\uparrow }\gg -\right. 
\end{equation}
\[
\left. -\int _{0}^{\infty }d\varepsilon {\rm Im}\ll c_{-\mathbf{k}_{1},\downarrow }|c_{\mathbf{k}_{2},\uparrow }\gg \int _{0}^{\infty }d\varepsilon {\rm Im}\ll c_{-\mathbf{k}_{2}+\mathbf{q},\downarrow }|c_{\mathbf{k}_{1}-\mathbf{q},\uparrow }\gg \right] .\]
Here, besides the before used SPGF's, the TPGF \( \ll c_{-\mathbf{k}_{1},\downarrow }c_{\mathbf{k}_{2},\uparrow }|c_{-\mathbf{k}_{2}+\mathbf{q},\downarrow }c_{\mathbf{k}_{1}-\mathbf{q},\uparrow }\gg  \)
appears. An explicit calculation of this function for non-perturbed (\( V_{{\rm L}}=0 \))
SC system (see Appendix E) gives the following result\begin{equation}
\label{2part}
\ll c_{-\mathbf{k}_{1},\downarrow }c_{\mathbf{k}_{2},\uparrow }|c_{-\mathbf{k}_{2}+\mathbf{q},\downarrow }c_{\mathbf{k}_{1}-\mathbf{q},\uparrow }\gg =\delta _{0,\mathbf{q}}\frac{\left( 4\varepsilon \Delta _{\mathbf{k}_{1}}\Delta _{\mathbf{k}_{2}}+\cdots \right) }{\left( \varepsilon ^{2}-E^{2}_{\mathbf{k}_{1}}-E^{2}_{\mathbf{k}_{2}}\right) ^{2}-4E^{2}_{\mathbf{k}_{1}}E^{2}_{\mathbf{k}_{2}}}.
\end{equation}
Here \( E_{\mathbf{k}}^{2}=\xi _{\mathbf{k}}^{2}+\Delta _{\mathbf{k}}^{2} \),
and three other terms in numerator are not mutually odd in \( \gamma _{\mathbf{k}_{1}} \)
and \( \gamma _{\mathbf{k}_{2}} \) and thus do not contribute into \( \delta ^{2} \).
Then, it is easy to see that the resulting contribution into Eq. \ref{disp2}
from M-diagonal (\( \mathbf{q}=0 \)) and N-non-diagonal TPGF's by Eq. \ref{2part}
exactly cancels with that from SPGF's, which confirms uniformity of \( d \)-wave
order in this case. The most important contribution to \( \delta ^{2} \) at
\( V_{{\rm L}}\neq 0 \) comes from two consecutive scattering processes in
the l.h.s. of an N- and M-non-diagonal (\( \mathbf{q}\neq 0 \)) TPGF, first
\( c_{-\mathbf{k}_{1},\downarrow }c_{\mathbf{k}_{2},\uparrow }\: \rightarrow \: c_{-\mathbf{k}_{1}+\mathbf{q},\downarrow }c_{\mathbf{k}_{2},\uparrow } \)
and then \( c_{-\mathbf{k}_{1}+\mathbf{q},\downarrow }c_{\mathbf{k}_{2},\uparrow }\: \rightarrow \: c_{-\mathbf{k}_{1}+\mathbf{q},\downarrow }c_{\mathbf{k}_{2}-\mathbf{q},\uparrow } \)
(or vice versa, Fig. \ref{fig2}), on the same scattering center \( \mathbf{p} \).
It is linear in \( c \), while the contribution from SPGF's in this order,
\( -2\Delta V\int _{0}^{\infty }d\varepsilon {\rm Im\: Tr}\widehat{\tau }_{1}\widehat{\Sigma } \),
is zero, accordingly to N-diagonal form of \( \widehat{\Sigma } \) established
in previous Section. Hence we generally estimate the gap dispersion to grow
with \( c \) as \( \left| \delta \right| =\Delta \sqrt{c/c_{1}} \), where
\( c_{1}\sim (V/V_{{\rm L}})^{2} \) defines the upper critical concentration
for SC at \( T=0 \).

A more detailed analysis, resolving amplitude and phase fluctuations, can be
done in a similar way, but considering separately two operators:\[
\Omega _{\mathbf{n},+}=\frac{\Omega _{\mathbf{n}}+\Omega _{\mathbf{n}}^{+}}{2},\quad {\rm and}\quad \Omega _{\mathbf{n},-}=\frac{\Omega _{\mathbf{n}}-\Omega _{\mathbf{n}}^{+}}{2i},\]
 such that their mean values \( \left\langle \Omega _{\mathbf{n},\pm }\right\rangle  \)
lead to real and imaginary parts of the order parameter, and constructing the
corresponding dispersions:\[
\delta _{\pm }^{2}=\frac{1}{N}\sum _{\mathbf{n}}\left( \left\langle \Omega _{\mathbf{n},\pm }^{2}\right\rangle -\left\langle \Omega _{\mathbf{n},\pm }\right\rangle ^{2}\right) .\]
This approach should be particularly important at extension of the theory to
finite temperatures, in order to establish the dominating type of fluctuations
due to the static disorder, responcible for breakdown of SC order at \( T=T_{c} \),
and its possible role in the persistence of a pseudogap in the density of states
\( \rho  \) at \( T>T_{c} \).

\section{Conclusions}

The above presented analysis shows that the disordered structure of doped HTSC
systems is crucial for many of their characteristic properties and for existence
of SC order itself. The interplay between doping and disorder effects can be
briefly resumed as follows. Superconductivity onsets with the metallization
of the system, at critical concentration \( c\sim c_{0} \) resulted from the
competition between kinetic energy of charge carriers in regular lattice and
their attraction to random dopant centers. The uniform \( d \)-wave order parameter
\( \Delta  \) develops with growing number of charge carriers as \( \Delta \sim \sqrt{c} \)
\cite{LokQuick} and saturates at certain optimum doping \( c_{opt}\sim \varepsilon _{{\rm D}}/W \),
when the relation \( \mu >\varepsilon _{{\rm D}} \) becomes to hold. With further
growing \( c \), the increasing local fluctuations of \( \Delta  \) bring
it to collapse at some upper critical concentration \( c_{1}\sim (V/V_{{\rm L}})^{2} \),
resulted from the competition between pairing and scattering potentials. This
picture is quantitatively satisfied with a very natural choice of parameters
\( W\sim 2 \) eV, \( V_{{\rm L}}\sim 0.5 \) eV, \( V\sim 0.22 \) eV, \( \varepsilon _{{\rm D}}\sim 0.2 \)
eV, giving plausible estimates: \( c_{0}\sim 5 \)\%, \( c_{opt}\sim 15 \)\%,
\( T_{c,max}\sim 100 \)K, \( c_{1}\sim 20 \)\%. The forthcoming work should
also specify such important moments, left beyond the scope of this consideration,
as the disorder effects on the cusp of density of states \( \rho \left( \varepsilon \right)  \)
at \( \varepsilon =\Delta  \), the matching conditions between the self-consistent
and GE descriptions of SPGF, the exact numerical coefficient for the critical
value \( c_{1} \), etc. And, of course, it is of principal interest to extend
the present self-consistent treatment to the case of finite temperatures up
to \( T_{c} \), in order to obtain a quantitative estimate for the bell-like
\( T_{c}(c) \) shape, and further to \( T>T_{c} \), to study the role of doping
disorder vs \( d \)-wave SC coupling in the formation and subsequent merging
(at \( c\sim c_{opt} \)) of the pseudogap in normal density of states.

This work was supported in parts by Portuguese Program PRAXIS XXI under the
Grant 2/2.1/FIS/302/94, by NATO Program OUTREACH under the Grant CP/UN/19/C/2000/PO,
and by Swiss National Science Foundation (SCOPES Project 7UKPJ062150.00/1).
One of us (V.M.L.) expresses his gratitude for a warm hospitality he felt at
visiting Centro de Física do Porto.

\section{Appendix A}

For the normal phase, we consider the dispersion law \( \varepsilon _{\mathbf{k}}=4t-2t\left( \cos ak_{x}+\cos ak_{y}\right)  \)
and relate the dopant concentration \( c \) to the number (per unit cell) of
occupied states below \( \mu  \):\begin{equation}
\label{A1}
c=\frac{2a^{2}}{\left( 2\pi \right) ^{2}}\int _{\varepsilon _{\mathbf{k}}\leq \mu }d\mathbf{k}=\frac{2a^{2}}{\pi ^{2}}\int ^{\arccos \left( 1-4\mu /W\right) }_{0}dk_{x}\int ^{\arccos \left( 2-4\mu /W-\cos ak_{x}\right) }_{0}dk_{y}=
\end{equation}
\[
=\frac{2}{\pi ^{2}}\int ^{1}_{1-4\mu /W}\frac{du}{\sqrt{1-u^{2}}}\int ^{1}_{2-4\mu /W-u}\frac{dv}{\sqrt{1-v^{2}}}=\frac{2}{\pi ^{2}}F\left( \frac{4\mu }{W}\right) .\]
 Here the dependence of the integral \[
F\left( x\right) =\int ^{1}_{1-x}\frac{\arccos \left( 2-x-u\right) du}{\sqrt{1-u^{2}}}\]
is very well approximated by a simple linear function \( F\left( x\right) \approx 5x/3 \)
in the whole physically important range \( 0\leq x=4\mu /W\leq x_{max} \) (where
\( x_{max}\approx 0.6 \) corresponds to \( \mu _{max}\approx 0.15W \) at maximum
physical doping \( c_{max}\approx 0.2 \)). Then we readily arrive at the estimate
\[
\mu \approx \frac{3\pi ^{2}}{40}cW\approx \frac{3cW}{4},\]
cited above.

\section{Appendix B}

For the uniform SC system with \( d \)-wave gap, we perform integration over
the Brillouin zone with the parametrization: \( \mathbf{k}-\mathbf{k}_{i}=a^{-1}(\rho _{0}\xi \mathbf{e}_{i}+\Delta ^{-1}\eta \mathbf{e}_{i}\times \mathbf{e}_{z}) \),
\( \mathbf{e}_{i}=\left( \pm 1/\sqrt{2},\pm 1/\sqrt{2},0\right)  \), \( \mathbf{e}_{z}=\left( 0,0,1\right)  \),
near 4 nodal points \( \mathbf{k}_{i}=\arccos \left( 1-\mu /W\right) \mathbf{e}_{i} \)
of the gap function \( \Delta _{\mathbf{k}}=\Delta \gamma ^{\left( d\right) }_{\mathbf{k}} \).
This integration for \( \widehat{G}^{\left( 0\right) } \)turns into\begin{equation}
\label{B1}
\widehat{G}^{\left( 0\right) }=\frac{1}{N}\sum _{\mathbf{k}}\widehat{G}^{\left( 0\right) }_{\mathbf{k}}=\rho _{0}\left[ \frac{1}{2\Delta }\int _{-\Delta }^{\Delta }d\eta \int _{-\varepsilon _{{\rm D}}}^{\varepsilon _{{\rm D}}}d\xi \widehat{g}\left( \varepsilon ,\xi ,\eta \right) +\right. 
\end{equation}
\[
\left. +\int _{-\mu }^{-\varepsilon _{{\rm D}}}d\xi \widehat{g}\left( \varepsilon ,\xi ,0\right) +\int _{\varepsilon _{{\rm D}}}^{W-\mu }d\xi \widehat{g}\left( \varepsilon ,\xi ,0\right) \right] ,\]
where we have defined the matrix function\[
\widehat{g}\left( \varepsilon ,\xi ,\eta \right) =\left( \varepsilon -\xi \widehat{\tau }_{3}-\eta \widehat{\tau }_{1}\right) ^{-1}=\frac{\varepsilon +\xi \widehat{\tau }_{3}+\eta \widehat{\tau }_{1}}{\varepsilon ^{2}-\xi ^{2}-\eta ^{2}}.\]
The integration in \( \xi  \) (normal to the Fermi surface) in Eq. \ref{B1}
treats the BCS shell, \( [-\varepsilon _{{\rm D}},\varepsilon _{{\rm D}}] \),
separately from the out-of-shell segments, \( [-\mu ,-\varepsilon _{{\rm D}}] \)
and \( [\varepsilon _{{\rm D}},W-\mu ] \), where the gap parameter \( \Delta  \)
turns zero (together with \( \gamma _{\mathbf{k}} \)) and no integration in
\( \eta =\Delta \gamma _{\mathbf{k}} \) is needed. Eq. \ref{B1} permits to
define explicitly the coefficient functions \( g_{i} \) in the general form
\( \widehat{G}^{\left( 0\right) }=g_{0}+g_{1}\widehat{\tau }_{1}+g_{3}\widehat{\tau }_{3} \).
Let us denote \( z^{2}=\varepsilon ^{2}-\eta ^{2} \), then the shell contribution
to \( g_{0} \) result from the integral:\begin{equation}
\label{B2}
\int _{-\varepsilon _{{\rm D}}}^{\varepsilon _{{\rm D}}}\frac{d\xi }{z^{2}-\xi ^{2}}=\frac{2}{z}\left( {\rm arctanh}\frac{z}{\varepsilon _{{\rm D}}}+i\pi \right) \approx \frac{2}{\varepsilon _{{\rm D}}}+\frac{2i\pi }{z},
\end{equation}
which is followed for its last term by:\begin{equation}
\label{B3}
\int _{-\Delta }^{\Delta }\frac{d\eta }{\sqrt{\varepsilon ^{2}-\eta ^{2}}}=2{\rm arccos}\frac{\Delta }{\varepsilon }.
\end{equation}
The out-of-shell contributions are:\begin{equation}
\label{B4}
\int _{-\mu }^{-\varepsilon _{{\rm D}}}\frac{d\xi }{\varepsilon ^{2}-\xi ^{2}}=\frac{1}{\varepsilon }\left( {\rm arctanh}\frac{\varepsilon }{\mu }-{\rm arctanh}\frac{\varepsilon }{\varepsilon _{{\rm D}}}\right) \approx \frac{1}{\mu }-\frac{1}{\varepsilon _{{\rm D}}}
\end{equation}
and\begin{equation}
\label{B5}
\int _{\varepsilon _{{\rm D}}}^{W-\mu }\frac{d\xi }{\varepsilon ^{2}-\xi ^{2}}\approx \frac{1}{W-\mu }-\frac{1}{\varepsilon _{{\rm D}}}.
\end{equation}
To find \( g_{1} \) and \( g_{3} \), we use the evident equalities \[
\int _{-\varepsilon _{{\rm D}}}^{\varepsilon _{{\rm D}}}\frac{\xi d\xi }{z^{2}-\xi ^{2}}=\int _{-1}^{1}\eta d\eta =0,\]
and:

\begin{equation}
\label{B6}
\int _{-\mu }^{-\varepsilon _{{\rm D}}}\frac{\xi d\xi }{\varepsilon ^{2}-\xi ^{2}}+\int _{\varepsilon _{{\rm D}}}^{W-\mu }\frac{\xi d\xi }{\varepsilon ^{2}-\xi ^{2}}=\frac{1}{2}\ln \frac{\mu ^{2}-\varepsilon ^{2}}{\left( W-\mu \right) ^{2}-\varepsilon ^{2}}\approx \ln \frac{\mu }{W-\mu }.
\end{equation}
Summing up Eqs. \ref{B2}-\ref{B6}, we obtain\[
g_{0}=\varepsilon \rho _{0}\left[ \frac{W}{\mu \left( W-\mu \right) }-\frac{\pi }{\Delta }\left( {\rm arccosh}\frac{\Delta }{\varepsilon }-\frac{i\pi }{2}\right) \right] \]
 in accordance with Eq. \ref{g0an}, \( g_{1}=0 \), and \( g_{3}=\rho _{0}\ln [\mu /(W-\mu )] \).

\section{Appendix C}

The gap equation, Eq. \ref{gap1}, for uniform (\( V_{{\rm L}}=0 \)) \( d \)-wave
system at \( T=0 \) transforms into: \begin{equation}
\label{C1}
\frac{1}{\lambda }=\frac{16}{\pi \Delta ^{3}}{\rm Im}\int _{0}^{\Delta }\eta ^{2}d\eta \int _{0}^{\varepsilon _{{\rm D}}}d\xi \int _{0}^{\infty }\frac{d\varepsilon }{\varepsilon ^{2}-\xi ^{2}-\eta ^{2}-i0},
\end{equation}
with \( \lambda =V\rho _{0} \). The sought quantity is the gap amplitude \( \Delta  \).
On the r.h.s. of Eq. \ref{C1} we perform the elementary integration in \( \varepsilon  \),
using the relation \( {\rm Im}\left( x-i0\right) ^{-1}=\pi \delta \left( x\right)  \):

\begin{equation}
\label{C2}
\frac{16}{\pi }{\rm Im}\int _{0}^{\Delta }\eta ^{2}d\eta \int _{0}^{\varepsilon _{{\rm D}}}d\xi \int _{0}^{\infty }\frac{d\varepsilon }{\varepsilon ^{2}-\xi ^{2}-\eta ^{2}-i0}=
\end{equation}
\[
=8\int _{0}^{\Delta }\eta ^{2}d\eta \int _{0}^{\varepsilon _{{\rm D}}}\frac{d\xi }{\sqrt{\xi ^{2}+\eta ^{2}}}\]
Then, integrating out in \( \xi  \) and passing from \( \eta  \) to \( y=\hbar \omega _{D}/\eta  \),
we present Eq. \ref{C2} as

\[
8\left( \frac{\varepsilon _{{\rm D}}}{\Delta }\right) ^{3}\int _{\frac{\varepsilon _{{\rm D}}}{\Delta }}^{\infty }\frac{{\rm arcsinh}y\: dy}{y^{4}}=\frac{8}{3}{\rm arcsinh}\frac{\varepsilon _{{\rm D}}}{\Delta }+4\frac{\varepsilon _{{\rm D}}\sqrt{\varepsilon _{{\rm D}}^{2}+\Delta ^{2}}}{3\Delta ^{2}}\]
\[
-\frac{4}{3}\left( \frac{\varepsilon _{{\rm D}}}{\Delta }\right) ^{3}{\rm arcsinh}\frac{\Delta }{\varepsilon _{{\rm D}}}\approx \frac{8}{3}\left( \ln \frac{2\varepsilon _{{\rm D}}}{\Delta }+\frac{1}{3}\right) .\]
 The equation for critical temperature \( T_{c} \), corresponding to \( \Delta =0 \),
reads in this case\[
\frac{1}{\lambda }=\frac{16}{3\pi }{\rm Im}\int _{0}^{\varepsilon _{{\rm D}}}d\xi \tanh \left( \frac{\xi }{2k_{B}T_{c}}\right) \int _{0}^{\infty }\frac{d\varepsilon }{\varepsilon ^{2}-\xi ^{2}-i0}\approx \frac{8}{3}\ln \left( \frac{2\gamma _{{\rm E}}\varepsilon _{{\rm D}}}{\pi k_{B}T_{c}}\right) ,\]
with the Euler constant \( \gamma _{{\rm E}}\approx 1.781 \). Hence for the
\( d \)-wave case the effective coupling constant is \( \widetilde{\lambda }=8\lambda /3 \),
that is \( 8/3 \) times the {}``Hamiltonian{}'' value \( \lambda  \), which
can serve as one more explanation for high \( T_{c} \) itself. Also, the ratio
\( r=2\Delta /k_{B}T_{c} \) results here \( \exp \left( 1/3\right)  \) times
the common \( s \)-wave BCS value \( r_{s}=2\pi /\gamma _{{\rm E}}\approx 3.52 \),
reaching as high value as \( r_{d}\approx 4.92 \). In its turn, this means
that, for equal other conditions (say, \( \rho _{0} \) and \( V \)), \( s \)-condensate
turns out more stable to thermal fluctuations and requires a higher \( T_{c} \)
to destroy it than \( d \)-condensate. Of course, this is directly related
to the absence of gap in the latter case, permitting quasi-particles to exist
at any \( T<T_{c} \).

\section{Appendix D}

Calculation of the self-consistent SPGF \( \widehat{G}^{\left( sc\right) }=G_{0}+G_{1}\widehat{\tau }_{1}+G_{3}\widehat{\tau }_{3} \)
generalizes the scheme of Appendix B:

\begin{equation}
\label{D1}
\widehat{G}^{\left( sc\right) }=\frac{1}{N}\sum _{\mathbf{k}}\widehat{G}^{\left( sc\right) }_{\mathbf{k}}=\rho _{0}\left[ \frac{1}{2\Delta }\int _{-\Delta }^{\Delta }d\eta \int _{-\varepsilon _{{\rm D}}}^{\varepsilon _{{\rm D}}}d\xi \widehat{g}\left( \varepsilon -\Sigma _{0},\xi +\Sigma _{3},\eta +\Sigma _{1}\right) +\right. 
\end{equation}
\[
\left. +\int _{-\mu }^{-\varepsilon _{{\rm D}}}d\xi \widehat{g}\left( \varepsilon -\Sigma _{0},\xi +\Sigma _{3},\Sigma _{1}\right) +\int _{\varepsilon _{{\rm D}}}^{W-\mu }d\xi \widehat{g}\left( \varepsilon -\Sigma _{0},\xi +\Sigma _{3},\Sigma _{1}\right) \right] .\]
Next we set \( z^{2}=(\varepsilon -\Sigma _{0})^{2}-(\eta +\Sigma _{1})^{2} \)
and \( z=r+i\gamma  \), and pass from \( \xi  \) to \( x=\xi +\Sigma _{3} \),
where \( \Sigma _{3} \) can be taken real (as seen e.g. from the final result
Eq. \ref{sc5}). Then the analogue to Eq. \ref{B2} is:\[
\int _{-\varepsilon _{{\rm D}}+\Sigma _{3}}^{\varepsilon _{{\rm D}}+\Sigma _{3}}\frac{dx}{z^{2}-x^{2}}=\frac{1}{2z}\times \]
\[
\times \left\{ \frac{1}{2}\ln \frac{\left[ \left( \varepsilon _{{\rm D}}+\Sigma _{3}+r\right) ^{2}+\gamma ^{2}\right] \left[ \left( \varepsilon _{{\rm D}}-\Sigma _{3}+r\right) ^{2}+\gamma ^{2}\right] }{\left[ \left( \varepsilon _{{\rm D}}-\Sigma _{3}-r\right) ^{2}+\gamma ^{2}\right] \left[ \left( \varepsilon _{{\rm D}}+\Sigma _{3}-r\right) ^{2}+\gamma ^{2}\right] }-\right. \]
\[
+i\left[ {\rm arctan}\frac{\varepsilon _{{\rm D}}+\Sigma _{3}+r}{\gamma }+{\rm arctan}\frac{\varepsilon _{{\rm D}}+\Sigma _{3}-r}{\gamma }+\right. \]
\[
\left. \left. +{\rm arctan}\frac{\varepsilon _{{\rm D}}-\Sigma _{3}+r}{\gamma }+{\rm arctan}\frac{\varepsilon _{{\rm D}}-\Sigma _{3}-r}{\gamma }\right] \right\} \approx \]
\begin{equation}
\label{D2}
\approx \frac{2}{\varepsilon _{{\rm D}}}-i\frac{\pi +2\gamma /\varepsilon _{{\rm D}}}{z},
\end{equation}
 where the small term \( 2\gamma /\varepsilon _{{\rm D}} \) can be safely dropped.
The next integration, in \( y=\eta +\Sigma _{1} \), is done only on \( i\pi /z \)
term accordingly to:\begin{equation}
\label{D3}
\int _{-\Delta +\Sigma _{1}}^{\Delta +\Sigma _{1}}\frac{dy}{\sqrt{(\varepsilon -\Sigma _{0})^{2}-y^{2}}}={\rm arccos}\frac{\Delta +\Sigma _{1}}{\varepsilon -\Sigma _{0}}+{\rm arccos}\frac{\Delta -\Sigma _{1}}{\varepsilon -\Sigma _{0}},
\end{equation}
which is relevant for \( G_{0} \), supplemented with\[
\int _{-\Delta +\Sigma _{1}}^{\Delta +\Sigma _{1}}\frac{ydy}{\sqrt{(\varepsilon -\Sigma _{0})^{2}-y^{2}}}=\]
\begin{equation}
\label{D4}
=-\frac{4\Delta \Sigma _{1}}{\sqrt{(\varepsilon -\Sigma _{0})^{2}-\left( \Delta +\Sigma _{1}\right) ^{2}}+\sqrt{(\varepsilon -\Sigma _{0})^{2}-\left( \Delta -\Sigma _{1}\right) ^{2}}}
\end{equation}
for \( G_{1} \). The out-of-shell integration of the mentioned components is
much easier, giving:\[
\int _{\varepsilon _{{\rm D}}-\Sigma _{3}}^{\mu -\Sigma _{3}}\frac{dx}{(\varepsilon -\Sigma _{0})^{2}-x^{2}}=\frac{1}{\varepsilon -\Sigma _{0}}\left( {\rm arctanh}\frac{\varepsilon -\Sigma _{0}}{\mu -\Sigma _{3}}-{\rm arctanh}\frac{\varepsilon -\Sigma _{0}}{\varepsilon _{{\rm D}}-\Sigma _{3}}\right) \approx \]
\begin{equation}
\label{D5}
\approx \frac{1}{\mu -\Sigma _{3}}-\frac{1}{\varepsilon _{{\rm D}}-\Sigma _{3}},
\end{equation}
and\begin{equation}
\label{D6}
\int _{\varepsilon _{{\rm D}}-\Sigma _{3}}^{W-\mu -\Sigma _{3}}\frac{dx}{(\varepsilon -\Sigma _{0})^{2}-x^{2}}\approx \frac{1}{W-\mu -\Sigma _{3}}-\frac{1}{\varepsilon _{{\rm D}}-\Sigma _{3}}.
\end{equation}
Here also \( \Sigma _{3} \) can be disregarded beside \( W,\mu ,\varepsilon _{{\rm D}} \),
then the two \( -1/\varepsilon _{{\rm D}} \) terms cancel out with that from
Eq. \ref{D2}. Now, combining Eqs. \ref{D2}-\ref{D6}, we obtain\[
G_{0}=(\varepsilon -\Sigma _{0})\rho _{0}\left[ \frac{W}{\mu \left( W-\mu \right) }-\right. \]
 \begin{equation}
\label{D7}
\left. \frac{\pi }{2\Delta }\left( {\rm arccosh}\frac{\Delta +\Sigma _{1}}{\varepsilon -\Sigma _{0}}+{\rm arccosh}\frac{\Delta -\Sigma _{1}}{\varepsilon -\Sigma _{0}}-i\pi \right) \right] ,
\end{equation}
and\[
G_{1}=\Sigma _{1}\rho _{0}\left[ \frac{W}{\mu \left( W-\mu \right) }-\frac{2i\pi }{\sqrt{(\varepsilon -\Sigma _{0})^{2}-\left( \Delta +\Sigma _{1}\right) ^{2}}+\sqrt{(\varepsilon -\Sigma _{0})^{2}-\left( \Delta -\Sigma _{1}\right) ^{2}}}\right] .\]
At least, \( G_{3} \) is obtained after\[
\int _{-\varepsilon _{{\rm D}}+\Sigma _{3}}^{\varepsilon _{{\rm D}}+\Sigma _{3}}\frac{xdx}{z^{2}-x^{2}}=\frac{1}{2}\ln \frac{\left( \varepsilon _{{\rm D}}-\Sigma _{3}\right) ^{2}-z^{2}}{\left( \varepsilon _{{\rm D}}+\Sigma _{3}\right) ^{2}-z^{2}},\]
\[
\int _{\varepsilon _{{\rm D}}+\Sigma _{3}}^{W-\mu +\Sigma _{3}}\frac{xdx}{z^{2}-x^{2}}=\frac{1}{2}\ln \frac{\left( \varepsilon _{{\rm D}}+\Sigma _{3}\right) ^{2}-z^{2}}{\left( W-\mu +\Sigma _{3}\right) ^{2}-z^{2}},\]
\[
\int _{-\mu +\Sigma _{3}}^{-\varepsilon _{{\rm D}}+\Sigma _{3}}\frac{xdx}{z^{2}-x^{2}}=\frac{1}{2}\ln \frac{\left( \mu -\Sigma _{3}\right) ^{2}-z^{2}}{\left( \varepsilon _{{\rm D}}-\Sigma _{3}\right) ^{2}-z^{2}},\]
\[
\frac{1}{2\Delta }\int _{-\Delta +\Sigma _{1}}^{\Delta +\Sigma _{1}}y^{2}dy=\frac{\Delta ^{2}}{3}+\Sigma _{1}^{2},\]
in the form\begin{equation}
\label{D8}
G_{3}=\rho _{0}\left\{ \ln \frac{\mu }{W-\mu }+2\Sigma _{3}\frac{(\varepsilon -\sigma _{0})^{2}-\Delta ^{2}/3-\Sigma _{1}^{2}}{\varepsilon _{{\rm D}}^{3}}\right\} .
\end{equation}

\section{Appendix E}

Search for solutions of Eq. \ref{dys} in the form of GE consists in consecutive
iterations of its r.h.s., separating systematically the GF's already present
in previous iterations \cite{Iv}. Let us start from M-diagonal SPGF \( \widehat{G}_{\mathbf{k}} \),
than the iteration sequence begins with singling out the scattering term with
\( \widehat{G}_{\mathbf{k}} \) itself from those with \( \widehat{G}_{\mathbf{k}^{\prime },\mathbf{k}} \),
\( \mathbf{k}^{\prime }\neq \mathbf{k} \):\[
\widehat{G}_{\mathbf{k}}=\widehat{G}_{\mathbf{k}}^{\left( 0\right) }+\widehat{G}_{\mathbf{k}}^{\left( 0\right) }\frac{1}{N}\sum _{\mathbf{k}^{\prime },\mathbf{p}}{\rm e}^{i\left( \mathbf{k}-\mathbf{k}^{\prime }\right) \cdot \mathbf{p}}\widehat{V}\widehat{G}_{\mathbf{k}^{\prime },\mathbf{k}}=\]

\begin{equation}
\label{c1}
=\widehat{G}_{\mathbf{k}}^{\left( 0\right) }+c\widehat{G}_{\mathbf{k}}^{\left( 0\right) }\widehat{V}\widehat{G}_{\mathbf{k}}+\widehat{G}_{\mathbf{k}}^{\left( 0\right) }\widehat{V}\frac{1}{N}\sum _{\mathbf{k}^{\prime }\neq \mathbf{k},\mathbf{p}}{\rm e}^{i\left( \mathbf{k}-\mathbf{k}^{\prime }\right) \cdot \mathbf{p}}\widehat{G}_{\mathbf{k}^{\prime },\mathbf{k}}.
\end{equation}
Then for each \( \widehat{G}_{\mathbf{k}^{\prime },\mathbf{k}} \) we write
down again Eq. \ref{dys} and separate the scattering terms with \( \widehat{G}_{\mathbf{k}} \)
and \( \widehat{G}_{\mathbf{k}^{\prime },\mathbf{k}} \) in its r.h.s: \[
\widehat{G}_{\mathbf{k}^{\prime },\mathbf{k}}=\widehat{G}_{\mathbf{k}^{\prime }}^{\left( 0\right) }\widehat{V}\frac{1}{N}\sum _{\mathbf{k}^{\prime \prime },\mathbf{p}^{\prime }}{\rm e}^{i\left( \mathbf{k}^{\prime }-\mathbf{k}^{\prime \prime }\right) \cdot \mathbf{p}}\widehat{G}_{\mathbf{k}^{\prime \prime },\mathbf{k}}=\]
\[
=c\widehat{G}_{\mathbf{k}^{\prime }}^{\left( 0\right) }\widehat{V}\widehat{G}_{\mathbf{k}^{\prime },\mathbf{k}}+\widehat{G}_{\mathbf{k}^{\prime }}^{\left( 0\right) }\widehat{V}\frac{1}{N}{\rm e}^{i\left( \mathbf{k}^{\prime }-\mathbf{k}\right) \cdot \mathbf{p}}\widehat{G}_{\mathbf{k}}+\widehat{G}_{\mathbf{k}^{\prime }}^{\left( 0\right) }\widehat{V}\frac{1}{N}\sum _{\mathbf{p}^{\prime }\neq \mathbf{p}}{\rm e}^{i\left( \mathbf{k}^{\prime }-\mathbf{k}\right) \cdot \mathbf{p}^{\prime }}\widehat{G}_{\mathbf{k}}+\]
\begin{equation}
\label{c2}
+\widehat{G}_{\mathbf{k}^{\prime }}^{\left( 0\right) }\widehat{V}\frac{1}{N}\sum _{\mathbf{k}^{\prime \prime }\neq \mathbf{k},\mathbf{k}^{\prime };\mathbf{p}^{\prime }}{\rm e}^{i\left( \mathbf{k}^{\prime }-\mathbf{k}^{\prime \prime }\right) \cdot \mathbf{p}^{\prime }}\widehat{G}_{\mathbf{k}^{\prime \prime },\mathbf{k}}.
\end{equation}
Note that the \( \mathbf{p}^{\prime }=\mathbf{p} \) term which gives the phase
factor \( {\rm e}^{i\left( \mathbf{k}^{\prime }-\mathbf{k}\right) \cdot \mathbf{p}} \)
in the r.h.s. of Eq. \ref{c2}, coherent to that already figured in the last
sum in Eq. \ref{c1}, is explicitly separated from incoherent ones, \( {\rm e}^{i\left( \mathbf{k}^{\prime }-\mathbf{k}\right) \cdot \mathbf{p}^{\prime }} \),
\( \mathbf{p}^{\prime }\neq \mathbf{p} \) (but there will be no such separation
when doing 1st iteration of Eq. \ref{dys} for M-non-diagonal SPGF \( \widehat{G}_{\mathbf{k}^{\prime },\mathbf{k}} \)
itself). Continuing the sequence, we shall explicitly collect the terms with
the initial function \( \widehat{G}_{\mathbf{k}} \), resulted from: i) all
multiple scatterings on the same site \( \mathbf{p} \) and ii) on the same
pair of sites \( \mathbf{p} \) and \( \mathbf{p}^{\prime }\neq \mathbf{p} \).
Then summing of i) in \( \mathbf{p} \) produces the first term of GE, and,
if the pair processes are neglected, it will coincide with the well known CPA
result \cite{Ell}. The second term of GE, obtained by summing of ii) in \( \mathbf{p},\mathbf{p}^{\prime }\neq \mathbf{p} \),
contains interaction matrices \( \widehat{A}_{\mathbf{p}^{\prime },\mathbf{p}} \)
generated by multiply scattered functions \( \widehat{G}_{\mathbf{k}^{\prime },\mathbf{k}} \),
\( \mathbf{k}^{\prime }\neq \mathbf{k} \), etc., (including their own renormalization).
For instance, the iteration for a function \( \widehat{G}_{\mathbf{k}^{\prime \prime },\mathbf{k}} \)
with \( \mathbf{k}^{\prime \prime }\neq \mathbf{k},\mathbf{k}^{\prime } \)
in the last term in Eq. \ref{c2} will give:\[
\widehat{G}_{\mathbf{k}^{\prime \prime },\mathbf{k}}=\widehat{G}_{\mathbf{k}^{\prime \prime }}^{\left( 0\right) }\widehat{V}\frac{1}{N}\sum _{\mathbf{k}^{\prime \prime \prime },\mathbf{p}^{\prime \prime }}{\rm e}^{i\left( \mathbf{k}^{\prime \prime }-\mathbf{k}^{\prime \prime \prime }\right) \cdot \mathbf{p}^{\prime \prime }}\widehat{G}_{\mathbf{k}^{\prime \prime \prime },\mathbf{k}}=\]
\[
=\widehat{G}_{\mathbf{k}^{\prime \prime }}^{\left( 0\right) }\widehat{V}\frac{1}{N}{\rm e}^{i\left( \mathbf{k}^{\prime \prime }-\mathbf{k}\right) \cdot \mathbf{p}}\widehat{G}_{\mathbf{k}}+\widehat{G}_{\mathbf{k}^{\prime \prime }}^{\left( 0\right) }\widehat{V}\frac{1}{N}{\rm e}^{i\left( \mathbf{k}^{\prime \prime }-\mathbf{k}\right) \cdot \mathbf{p}^{\prime }}\widehat{G}_{\mathbf{k}}+\]
\begin{equation}
\label{c3}
+\; {\rm terms\; with}\; \widehat{G}_{\mathbf{k}^{\prime },\mathbf{k}}\; {\rm and}\; \widehat{G}_{\mathbf{k}^{\prime \prime },\mathbf{k}}\; +{\rm terms\; with}\; \widehat{G}_{\mathbf{k}^{\prime \prime \prime },\mathbf{k}}\; (\mathbf{k}^{\prime \prime \prime }\neq \mathbf{k},\mathbf{k}^{\prime },\mathbf{k}^{\prime \prime }).
\end{equation}
Consequently, the GE for \( \widehat{G}_{\mathbf{k}} \) obtains the form given
by Eq. \ref{sol}. 

Now turn to TPGF \( \ll c_{-\mathbf{k}_{1},\downarrow }c_{\mathbf{k}_{2},\uparrow }|c_{-\mathbf{k}_{2}+\mathbf{q},\downarrow }c_{\mathbf{k}_{1}-\mathbf{q},\uparrow }\gg  \),
beginning from EM's in absence of scattering which develop into a \( 4\times 4 \)
matrix form: \begin{equation}
\label{c4}
\widehat{B}\left( \xi _{1},\xi _{2},\Delta _{1},\Delta _{2}\right) f=\delta _{0,\mathbf{q}}d,
\end{equation}
\[
\widehat{B}\left( \xi _{1},\xi _{2},\Delta _{1},\Delta _{2}\right) =\left( \begin{array}{cccc}
\varepsilon -\xi _{1}-\xi _{2} & \Delta _{1} & \Delta _{2} & 0\\
\Delta _{1} & \varepsilon +\xi _{1}-\xi _{2} & 0 & \Delta _{2}\\
\Delta _{2} & 0 & \varepsilon -\xi _{1}+\xi _{2} & \Delta _{1}\\
0 & \Delta _{2} & \Delta _{1} & \varepsilon +\xi _{1}+\xi _{2}
\end{array}\right) ,\]
with 4-vectors: \[
f=\left( \begin{array}{c}
\ll c_{-\mathbf{k}_{1},\downarrow }c_{\mathbf{k}_{2},\uparrow }|c_{-\mathbf{k}_{2}+\mathbf{q},\downarrow }c_{\mathbf{k}_{1}-\mathbf{q},\uparrow }\gg \\
\ll c^{+}_{\mathbf{k}_{1},\uparrow }c_{\mathbf{k}_{2},\uparrow }|c_{-\mathbf{k}_{2}+\mathbf{q},\downarrow }c_{\mathbf{k}_{1}-\mathbf{q},\uparrow }\gg \\
\ll c_{-\mathbf{k}_{1},\downarrow }c^{+}_{-\mathbf{k}_{2},\downarrow }|c_{-\mathbf{k}_{2}+\mathbf{q},\downarrow }c_{\mathbf{k}_{1}-\mathbf{q},\uparrow }\gg \\
\ll c^{+}_{\mathbf{k}_{1},\uparrow }c^{+}_{-\mathbf{k}_{2},\downarrow }|c_{-\mathbf{k}_{2}+\mathbf{q},\downarrow }c_{\mathbf{k}_{1}-\mathbf{q},\uparrow }\gg 
\end{array}\right) ,\quad d=\left( \begin{array}{c}
2\Delta _{1}\Delta _{2}\\
\Delta _{1}\\
\Delta _{2}\\
2
\end{array}\right) ,\]
and \( \xi _{1}\equiv \xi _{\mathbf{k}_{1}} \), \( \xi _{2}\equiv \xi _{\mathbf{k}_{2}} \),
\( \Delta _{1}\equiv \Delta _{\mathbf{k}_{1}} \), \( \Delta _{2}\equiv \Delta _{\mathbf{k}_{2}} \).
The solution to Eq. \ref{c4}: \( f=\delta _{0,\mathbf{q}}\widehat{B}^{-1}\left( \xi _{1},\xi _{2},\Delta _{1},\Delta _{2}\right) d \),
gives just the result of Eq. \ref{2part} for the 1st component of \( f \). 

In presence of scattering, we consider only the M-non-diagonal (\( \mathbf{q}\neq 0 \))
case, then Eq. \ref{c4} turns into:\begin{equation}
\label{c5}
\widehat{B}\left( \xi _{1},\xi _{2},\Delta _{1},\Delta _{2}\right) f=-\frac{1}{N}\sum _{\mathbf{p}}\left( {\rm e}^{-i\mathbf{q}\cdot \mathbf{p}}\widehat{A}_{1}f^{\prime }_{1}+{\rm e}^{i\mathbf{q}\cdot \mathbf{p}}\widehat{A}_{2}f^{\prime }_{2}\right) ,
\end{equation}
where the vectors of {}``single scattered{}'' TPGF's are\[
f^{\prime }_{1}=\left( \begin{array}{c}
\ll c_{-\mathbf{k}_{1}+\mathbf{q},\downarrow }c_{\mathbf{k}_{2},\uparrow }|c_{-\mathbf{k}_{2}+\mathbf{q},\downarrow }c_{\mathbf{k}_{1}-\mathbf{q},\uparrow }\gg \\
\ll c^{+}_{\mathbf{k}_{1}-\mathbf{q},\uparrow }c_{\mathbf{k}_{2},\uparrow }|c_{-\mathbf{k}_{2}+\mathbf{q},\downarrow }c_{\mathbf{k}_{1}-\mathbf{q},\uparrow }\gg \\
\ll c_{-\mathbf{k}_{1}+\mathbf{q},\downarrow }c^{+}_{-\mathbf{k}_{2},\downarrow }|c_{-\mathbf{k}_{2}+\mathbf{q},\downarrow }c_{\mathbf{k}_{1}-\mathbf{q},\uparrow }\gg \\
\ll c^{+}_{\mathbf{k}_{1}-\mathbf{q},\uparrow }c^{+}_{-\mathbf{k}_{2},\downarrow }|c_{-\mathbf{k}_{2}+\mathbf{q},\downarrow }c_{\mathbf{k}_{1}-\mathbf{q},\uparrow }\gg 
\end{array}\right) ,\]
\[
f^{\prime }_{2}=\left( \begin{array}{c}
\ll c_{-\mathbf{k}_{1},\downarrow }c_{\mathbf{k}_{2}-\mathbf{q},\uparrow }|c_{-\mathbf{k}_{2}+\mathbf{q},\downarrow }c_{\mathbf{k}_{1}-\mathbf{q},\uparrow }\gg \\
\ll c^{+}_{\mathbf{k}_{1},\uparrow }c_{\mathbf{k}_{2}-\mathbf{q},\uparrow }|c_{-\mathbf{k}_{2}+\mathbf{q},\downarrow }c_{\mathbf{k}_{1}-\mathbf{q},\uparrow }\gg \\
\ll c_{-\mathbf{k}_{1},\downarrow }c^{+}_{-\mathbf{k}_{2}+\mathbf{q},\downarrow }|c_{-\mathbf{k}_{2}+\mathbf{q},\downarrow }c_{\mathbf{k}_{1}-\mathbf{q},\uparrow }\gg \\
\ll c^{+}_{\mathbf{k}_{1},\uparrow }c^{+}_{-\mathbf{k}_{2}+\mathbf{q},\downarrow }|c_{-\mathbf{k}_{2}+\mathbf{q},\downarrow }c_{\mathbf{k}_{1}-\mathbf{q},\uparrow }\gg 
\end{array}\right) ,\]
 and the \( 4\times 4 \) matrices:\[
\widehat{A}_{1}=\left( \begin{array}{cc}
\widehat{V} & 0\\
0 & \widehat{V}
\end{array}\right) ,\qquad \widehat{A}_{2}=\left( \begin{array}{cc}
\widehat{V}\widehat{\tau }_{3} & 0\\
0 & -\widehat{V}\widehat{\tau }_{3}
\end{array}\right) .\]
Next, EM's for \( f^{\prime }_{1,2} \):\begin{equation}
\label{c7}
\widehat{B}\left( \xi _{1}^{\prime },\xi _{2},\Delta ^{\prime }_{1},\Delta _{2}\right) f^{\prime }_{1}=-\frac{1}{N}{\rm e}^{-i\mathbf{q}\cdot \mathbf{p}}\widehat{A}_{1}f^{\prime \prime },
\end{equation}
\[
\widehat{B}\left( \xi _{1},\xi _{2}^{\prime },\Delta _{1},\Delta ^{\prime }_{2}\right) f^{\prime }_{2}=-\frac{1}{N}{\rm e}^{-i\mathbf{q}\cdot \mathbf{p}}\widehat{A}_{2}f^{\prime \prime },\]
with \( \xi _{1}^{\prime }\equiv \xi _{\mathbf{k}_{1}-\mathbf{q}} \), \( \xi _{2}^{\prime }\equiv \xi _{\mathbf{k}_{2}-\mathbf{q}} \),
\( \Delta ^{\prime }_{1}\equiv \Delta _{\mathbf{k}_{1}-\mathbf{q}} \), \( \Delta ^{\prime }_{2}\equiv \Delta _{\mathbf{k}_{2}-\mathbf{q}} \),
contain the {}``double scattered{}'' TPGF \( f^{\prime \prime }\equiv \ll c_{-\mathbf{k}_{1}+\mathbf{q},\downarrow }c_{\mathbf{k}_{2}-\mathbf{q},\uparrow }|c_{-\mathbf{k}_{2}+\mathbf{q},\downarrow }c_{\mathbf{k}_{1}-\mathbf{q},\uparrow }\gg  \),
which is already M-diagonal and hence can be taken just in the form of Eq. \ref{2part}.
Finally, the solution\begin{equation}
\label{c8}
f=\widehat{B}^{-1}\left( \xi _{1},\xi _{2},\Delta _{1},\Delta _{2}\right) \left\{ \delta _{0,\mathbf{q}}+\frac{c}{N}\left[ \widehat{A}_{1}\widehat{B}^{-1}\left( \xi _{1}^{\prime },\xi _{2},\Delta ^{\prime }_{1},\Delta _{2}\right) \widehat{A}_{1}+\right. \right. 
\end{equation}
\[
\left. \left. +\widehat{A}_{2}\widehat{B}^{-1}\left( \xi _{1},\xi _{2}^{\prime },\Delta _{1},\Delta ^{\prime }_{2}\right) \widehat{A}_{2}\right] \widehat{B}^{-1}\left( \xi _{1}^{\prime },\xi _{2}^{\prime },\Delta ^{\prime }_{1},\Delta ^{\prime }_{2}\right) \right\} d,\]
defines the contribution \( \sim c(V_{{\rm L}}\Delta /V)^{2} \) into \( \delta ^{2} \),
Eq. \ref{disp2}, the factor \( \Delta ^{2} \) being due to \( \Delta _{1},\Delta _{2} \)-odd
terms from \( \widehat{B}^{-1}\left( \xi _{1},\xi _{2},\Delta _{1},\Delta _{2}\right)  \)
and \( 1/V^{2} \) due to dominating, zeroth order in \( \Delta _{1},\Delta _{2} \),
terms from \( \widehat{B}^{-1}\left( \xi _{1}^{\prime },\xi _{2},\Delta ^{\prime }_{1},\Delta _{2}\right) \widehat{B}^{-1}\left( \xi _{1}^{\prime },\xi _{2}^{\prime },\Delta ^{\prime }_{1},\Delta ^{\prime }_{2}\right)  \)
and \( \widehat{B}^{-1}\left( \xi _{1},\xi _{2}^{\prime },\Delta _{1},\Delta ^{\prime }_{2}\right) \widehat{B}^{-1}\left( \xi _{1}^{\prime },\xi _{2}^{\prime },\Delta ^{\prime }_{1},\Delta ^{\prime }_{2}\right)  \).

\end{document}